\documentclass[fleqn,10pt]{wlscirep}
\usepackage[utf8]{inputenc}
\usepackage[T1]{fontenc}
\title{Spin effect on the low-temperature resistivity maximum in a strongly interacting 2D electron system}
\author[1]{A.~A. Shashkin}
\author[1]{M.~Yu.\ Melnikov}
\author[1]{V.~T. Dolgopolov}
\author[2]{M.~M. Radonji\'c}
\author[3]{V. Dobrosavljevi\'c}
\author[4]{S.-H. Huang}
\author[4]{C.~W. Liu}
\author[5]{Amy~Y.~X.~Zhu}
\author[5,*]{S.~V. Kravchenko}

\affil[1]{Institute of Solid State Physics, Chernogolovka, Moscow District 142432, Russia}
\affil[2]{Institute of Physics Belgrade, University of Belgrade, Pregrevica 118, 11080 Belgrade, Serbia}
\affil[3]{Department of Physics and National High Magnetic Field Laboratory, Florida State University, Tallahassee, Florida 32306, USA}
\affil[4]{Department of Electrical Engineering and Graduate Institute of Electronics Engineering, National Taiwan University, Taipei 106, Taiwan}
\affil[5]{Department of Physics, Northeastern University, Boston, Massachusetts 02115, USA}
\affil[*]{s.kravchenko@northeastern.edu}

\begin{abstract}
\textbf{The increase in the resistivity with decreasing temperature followed by a drop by more than one order of magnitude is observed on the metallic side near the zero-magnetic-field metal-insulator transition in a strongly interacting two-dimensional electron system in ultra-clean SiGe/Si/SiGe quantum wells. We find that the temperature $T_{\text{max}}$, at which the resistivity exhibits a maximum, is close to the renormalized Fermi temperature. However, rather than increasing along with the Fermi temperature, the value $T_{\text{max}}$ decreases appreciably for spinless electrons in spin-polarizing (parallel) magnetic fields. The observed behaviour of $T_{\text{max}}$ cannot be described by existing theories. The results indicate the spin-related origin of the effect.
}
\end{abstract}

\begin{document}
\flushbottom
\maketitle

\thispagestyle{empty}

The zero-magnetic-field metal-insulator transition has been reported in a number of strongly-correlated two-dimensional (2D) electron systems in semiconductors \cite{kravchenko1994possible,popovic1997metal,coleridge1997metal,hanein1998the,hanein1998observation,papadakis1998apparent,gao2002weak,kravchenko2004metal,shashkin2005metal,pudalov2006metal,kozuka2014challenges,melnikov2019quantum,shashkin2019recent,shashkin2021metal}, in quasi-2D organic charge-transfer salts (Mott organics) \cite{dressel2020molecular}, as well as in 2D transition metal dichalcogenides \cite{moon2020quantum,moon2021metal,li2021continuous}. The hallmark of the low-temperature resistivity $\rho$ on the metallic side near the metal-insulator transition is a non-monotonic $\rho(T)$: when the temperature is decreased, the resistivity first increases, reaching a maximum at a temperature $T_{\text{max}}$, and then drops at lower temperatures. The strength of such a resistivity drop varies from system to system, reaching a 12-fold value in ultra-pure SiGe/Si/SiGe quantum wells \cite{melnikov2019quantum}; note that this is in contrast to the case of low-mobility Si/SiGe quantum wells where the resistivity decreases at low temperatures by a factor of about 1.5 and the disorder effects are dominant \cite{lai2004observation,lu2011termination}. The increase of the effective mass with decreasing electron density \cite{kravchenko2004metal,shashkin2005metal,shashkin2019recent,shashkin2021metal,shashkin2002sharp,mokashi2012critical,kuntsevich2015strongly,melnikov2017indication}
suggests that the effect of strong electron-electron interactions is the main driving force for the experimentally observed metal-insulator transition due to the fermion condensation \cite{amusia2015theory,melnikov2017indication} or the Wigner-Mott transition \cite{camjayi2008coulomb,brussarski2018transport}. From the theoretical standpoint, the dynamical mean-field theory (DMFT) approach \cite{camjayi2008coulomb,radonjic2012wigner,dobrosavljevic2017wigner} has provided a successful quantitative description of the low-temperature resistivity drop in zero magnetic field in all these systems, especially in SiGe/Si/SiGe quantum wells \cite{shashkin2020manifestation}. Over a wide range of electron densities, the experimental ratio $(\rho(T)-\rho(0))/(\rho(T_{\text{max}})-\rho(0))$ has been found to be a universal function of $T/T_{\text{max}}$, which is consistent with the DMFT. According to this theory, $T_{\text{max}}$ corresponds to the quasiparticle coherence temperature, which is of the order of the Fermi temperature $T_{\text F}$ determined by the effective electron mass renormalized by interactions: $T_{\text{max}}\sim T_{\text F}$. Below this temperature, the elastic electron-electron scattering corresponds to coherent transport, while at higher temperatures, the inelastic electron-electron scattering becomes strong and gives rise to a fully incoherent transport. Notably, similar functional form of the resistivity $\rho(T)$ can be expected within the screening theory in its general form (for more on this, see Ref.~\cite{shashkin2020manifestation}). Similar non-monotonic $\rho(T)$ dependence with a maximum at $T_{\text{max}}\sim T_{\text F}$ is predicted by another approach based on the Pomeranchuk effect expected within a phase coexistence region between the Wigner crystal and a Fermi liquid \cite{spivak2003phase,spivak2004phases,spivak2006transport}. Finally, the renormalization-group scaling theory, which takes account of the existence of multiple valleys in the electron spectrum \cite{punnoose2001dilute,punnoose2005metal}, allows for a quantitative description of non-monotonic $\rho(T)$ on the metallic side in the close vicinity of the metal-insulator transition for the resistivities low compared to $\pi h/e^2$ \cite{shashkin2020manifestation}. Within this theory, $T_{\text{max}}$ is determined by the competition between electron-electron interactions and disorder, the value of $T_{\text{max}}$ being much smaller than $T_{\text F}$ and decreasing when lifting the spin degeneracy.

The existence of the metallic state in strongly interacting 2D electron systems is intimately related to the existence of spin and valley degrees of freedom \cite{finkelstein1983influence,lee1985disordered,punnoose2001dilute,punnoose2005metal}. If the electron spins become completely polarized by a magnetic field $B^*$ parallel to the 2D plane, the spin degeneracy that determines the Fermi energy changes to $g_{\text s}=1$, corresponding to spinless electrons. In a thin 2D electron system in the metallic regime, the resistivity increases with parallel magnetic field by a factor of a few and saturates above the polarization field $B^*$ \cite{simonian1997magnetic,pudalov1997instability,okamoto1999spin,vitkalov2000small,yoon2000parallel,shashkin2001metal,vakili2004spin,gao2006spin}.  (An attempt was made by M.~S. Hossain \textit{et al}. \cite{hossain2020observation} to analyze the data for $B^*$ in the insulating phase. However, such data reflect the physics of localized electron moments in the band tail, which is entirely different from that of the metallic phase (see, \textit{e.g.}, Ref.~\cite{shashkin2005metal}), and, therefore, the conclusion of M.~S. Hossain \textit{et al}.\ on the ferromagnetic state is not justified.) The metallic temperature dependence of the resistivity is suppressed in the spin-polarized regime, as observed in silicon metal-oxide-semiconductor field-effect transistors (MOSFETs) \cite{simonian1997magnetic,shashkin2001metal,li2017resistivity}, $p$-type GaAs/AlGaAs heterostructures \cite{yoon2000parallel,gao2006spin}, and narrow AlAs quantum wells \cite{vakili2004spin}. Recent work has established that the ultra-clean 2D electron system in SiGe/Si/SiGe quantum wells, which is similar to the clean Si MOSFETs in that the 2D electrons host in Si but is distinguished mainly by the much higher electron mobility, still exhibits a metal-insulator transition even at $B=B^*$ that is determined using a number of different measurement methods \cite{shashkin2001metal,shashkin2005metal} and is attributed to the existence of two distinct valleys in its spectrum \cite{melnikov2020metallic}.

In this paper, we report studies of the non-monotonic temperature dependence of the resistivity on the metallic side near the metal-insulator transition in a strongly interacting, spin-unpolarized ($g_{\text s}=2$) as well as fully spin-polarized (or spinless, $g_{\text s}=1$) bi-valley 2D electron system in ultra-clean SiGe/Si/SiGe quantum wells. We find that in zero magnetic field, the temperature $T_{\text{max}}$, at which the resistivity has a maximum, is close to the renormalized Fermi temperature $T_{\text F}$, which is in agreement with the dynamical mean-field theory. However, rather than increasing along with the Fermi temperature, the value $T_{\text{max}}$ decreases appreciably for spinless electrons in spin-polarizing magnetic fields, which is in contradiction with this theory. A scaling analysis of $\rho(T)$ in the spinless electron system in the spirit of DMFT shows that the low-temperature resistivity drop is still described by the theory, similar to the case of the spin-unpolarized electron system. At the same time, the reduced value of $T_{\text{max}}$ in spin-polarizing magnetic fields is consistent with the predictions of the renormalization-group scaling theory, but $T_{\text{max}}$ in zero magnetic field is in disagreement with this theory. Thus, the observed behaviour of $T_{\text{max}}$ cannot be described by existing theories. Nor can it be explained in terms of the increase of the residual disorder potential and the reduction of the electron interaction strength due to the reduced spin degrees of freedom in spin-polarizing magnetic fields, because the relation $T_{\text{max}}\sim T_{\text F}$ still holds for clean Si MOSFETs \cite{radonjic2012wigner,dobrosavljevic2017wigner} and low-mobility Si/SiGe quantum wells \cite{lu2011termination} in zero magnetic field. This indicates the spin-related origin of the effect.

\begin{figure}
\scalebox{.48}{\includegraphics[angle=0]{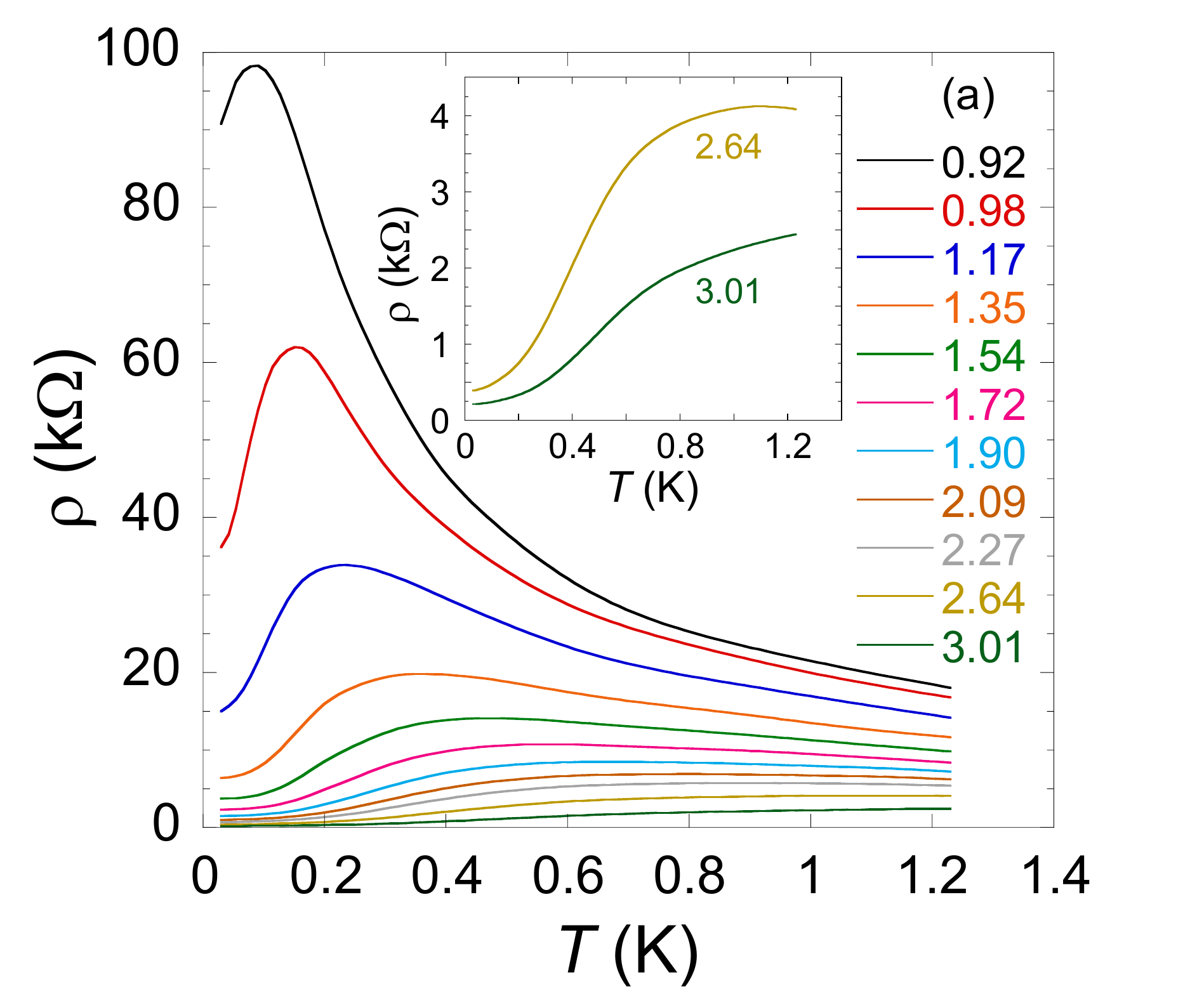}}
\scalebox{.48}{\includegraphics[angle=0]{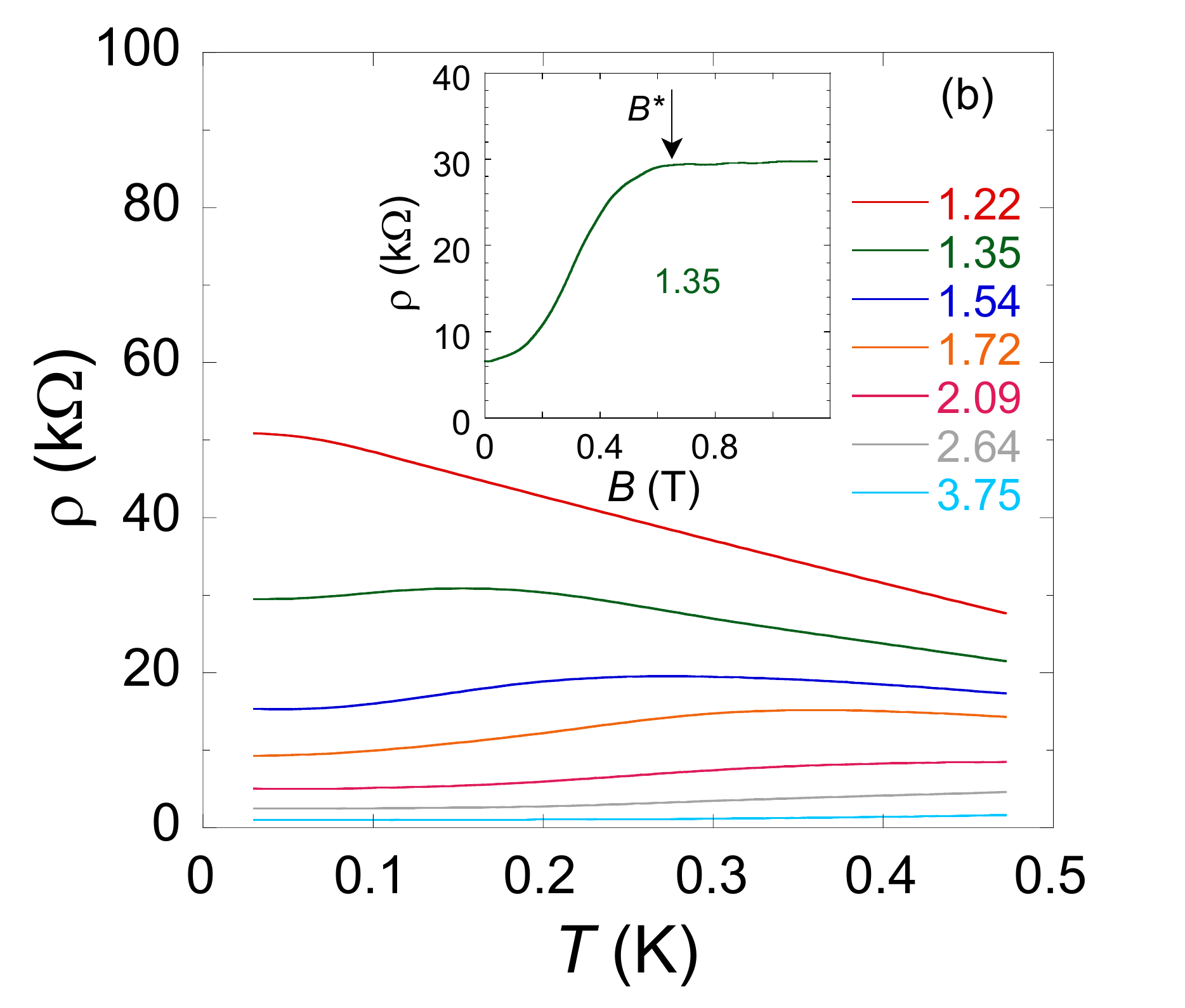}}
\caption{Non-monotonic temperature dependences of the resistivity of the 2D electron system in SiGe/Si/SiGe quantum wells on the metallic side near the metal-insulator transition (a) in $B=0$ and (b) in $B=B^*$; the magnetic fields used are spanned in the range between approximately 1 and 2 T. The electron densities are indicated in units of $10^{10}$~cm$^{-2}$. The inset in (a) shows $\rho(T)$ at $n_{\text s}=2.64$ and $3.01\times10^{10}$~cm$^{-2}$ on an expanded scale. The inset in (b) shows the parallel-field magnetoresistance at $n_{\text s}=1.35\times10^{10}$~cm$^{-2}$ at $T\approx 25$~mK. The polarization field $B^*$ is indicated.}\label{fig1} 
\end{figure}

\begin{figure}
\scalebox{0.48}{\includegraphics[angle=0]{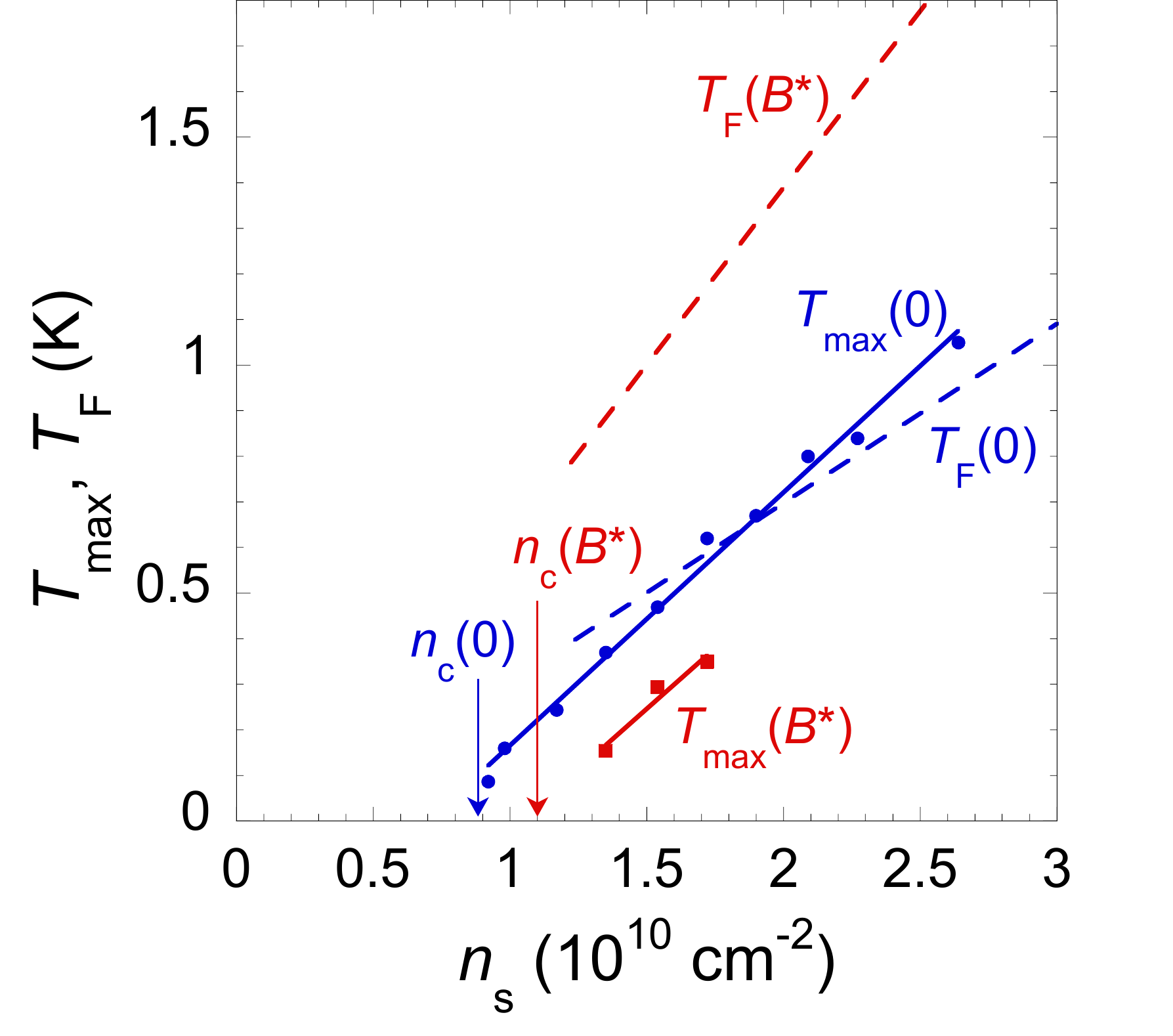}}
\caption{$T_{\text{max}}$ as a function of electron density in $B=0$ (circles) and in $B=B^*$ (squares). Solid lines are linear fits. Critical electron densities for the metal-insulator transition in $B=0$ and $B=B^*$ are indicated. Dashed lines show the Fermi temperatures $T_{\text F}$ in $B=0$ and $B=B^*$ calculated using the low-temperature value $B^*$ and Eq.~(\ref{eq1}), see text.}\label{fig2} 
\end{figure}

Raw data for the resistivity as a function of temperature are shown at zero magnetic field ($g_{\text s}=2$) in Fig.~\ref{fig1}(a) and at $B=B^*$ ($g_{\text s}=1$) in Fig.~\ref{fig1}(b) for electron densities above the critical electron densities for the metal-insulator transition $n_{\text c}(0)\approx 0.88\times10^{10}$~cm$^{-2}$ and $n_{\text c}(B^*)\approx 1.1\times10^{10}$~cm$^{-2}$, respectively. In zero magnetic field, the $\rho(T)$ curves are non-monotonic with the maxima at density-dependent temperatures $T_{\text{max}}(0)$ over a wide range of electron densities $n_{\text s}$; below $T_{\text{max}}(0)$, the resistivity drops sharply with decreasing temperature so that the drop can exceed an order of magnitude (see the inset to Fig.~\ref{fig1}(a)). At $B=B^*$, the $\rho(T)$ curves are non-monotonic in a narrower range of electron densities, and the resistivity drop below $T_{\text{max}}(B^*)$ weakens. The values of $B^*$ are density-dependent and have been determined by the saturation of the $\rho(B)$ dependences (see the inset to Fig.~\ref{fig1}(b)), which corresponds to the lifting of the spin degeneracy \cite{okamoto1999spin,vitkalov2000small}. Magnetic fields used in our experiments were within the range between approximately 1 and 2~T. The measurements were restricted to 0.5~K because this was the highest temperature at which the complete spin polarization could still be achieved; the restriction is likely to reflect the degeneracy condition for the dilute electron system with low Fermi energy.

\begin{figure}
\scalebox{0.48}{\includegraphics[angle=0]{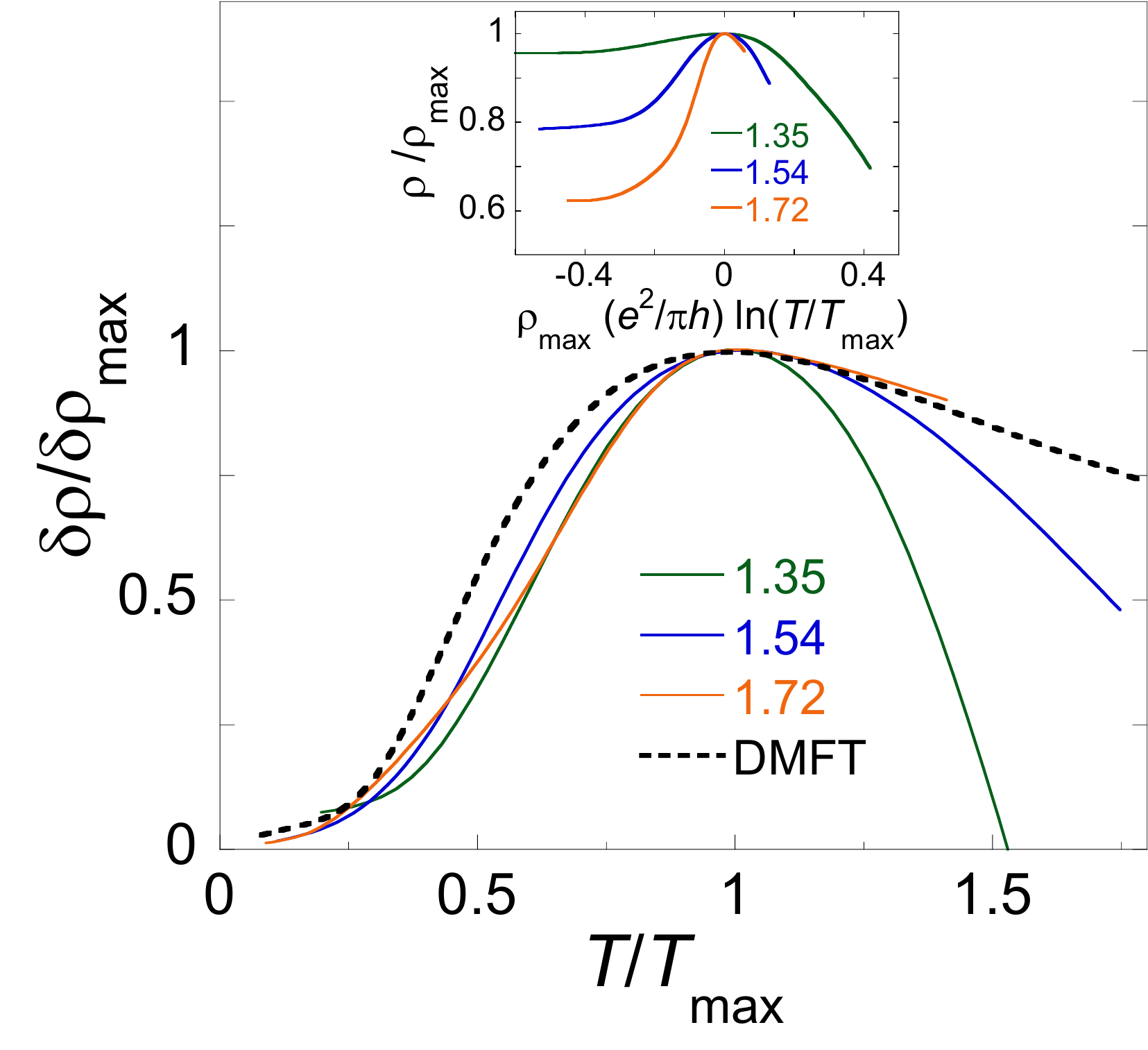}}
\caption{The ratio $\delta\rho/\delta\rho_{\text{max}}$ plotted as a function of $T/T_{\text{max}}$ in $B=B^*$. The electron densities are indicated in units of $10^{10}$~cm$^{-2}$. The dashed line is the result of DMFT in the weak-disorder limit \cite{camjayi2008coulomb,radonjic2012wigner,dobrosavljevic2017wigner}. The inset shows the analysis based on the scaling form suggested by the renormalization-group scaling theory \cite{punnoose2001dilute,punnoose2005metal}.}\label{fig3}
\end{figure}

In Fig.~\ref{fig2}, which is the main figure of this paper, we plot the values of $T_{\text{max}}$ as a function of the electron density in $B=0$ and $B=B^*$. The data for $T_{\text{max}}(B^*)$ lie significantly lower than those for $T_{\text{max}}(0)$. Interestingly, each dependence can be described by a linear function that extrapolates to zero at $n_{\text s}$ close to $n_{\text c}(0)$ or $n_{\text c}(B^*)$, and the slopes of both dependences are close to each other. We also plot the calculated values of renormalized Fermi temperatures $T_{\text F}$ for both cases. In zero magnetic field, the density dependences of the resistivity maximum temperature $T_{\text{max}}(0)$ and the Fermi temperature $T_{\text F}(0)$ are close to each other in the electron density range where they overlap. However, there is a qualitative difference between the behaviour of $T_{\text{max}}$ and that of $T_{\text F}$ when lifting the spin degeneracy. Rather than increasing along with the Fermi temperature, the value $T_{\text{max}}$ decreases when polarizing electron spins.

The Fermi temperature $T_{\text F}(B^*)$ has been calculated from the low-temperature value $B^*$ (see the inset to Fig.~\ref{fig1}(b)) based on the equality of the Fermi energy of completely spin-polarized electrons and the Zeeman energy in the polarization field $B^*$ \cite{melnikov2017indication}:
\begin{equation}
k_{\text B}T_{\text F}(B^*)=\frac{h^2n_{\text s}}{2\pi g_{\text v}m}=g_{\text F}\mu_{\text B}B^*,\label{eq1}
\end{equation}
where $k_{\text B}$ is the Boltzmann constant, $g_{\text v}=2$ is the valley degeneracy, $m$ is the renormalized energy-averaged effective mass that is determined by the density of states, $g_{\text F}\approx g_0=2$ is the $g$-factor at the Fermi level, $g_0$ is the $g$-factor in bulk silicon, and $\mu_{\text B}$ is the Bohr magneton. We argue that the Fermi temperature $T_{\text F}(0)$ of spin-unpolarized electrons is approximately half of the Fermi temperature $T_{\text F}(B^*)$ of completely spin-polarized ones. Indeed, it was experimentally shown in Ref.~\cite{shashkin2006pauli} that the electron spin magnetization is proportional to the parallel magnetic field in the range up to $B=B^*$ for the clean, strongly interacting 2D electron system in Si MOSFETs that is similar to the 2D electron system in SiGe/Si/SiGe quantum wells.  (For strongly disordered Si MOSFETs, the band tail of localized electrons persists into the metallic regime \cite{prus2003thermodynamic} in which case both the nonlinear magnetization as a function of parallel magnetic field and the shift of the dependence $B^*(n_{\text s})$ to higher densities are observed due to the presence of localized electron moments in the band tail \cite{dolgopolov2002comment,gold2002on,shashkin2005metal,teneh2012spin,pudalov2018probing}.) Taking into account the smallness of the exchange effects in the 2D electron system in silicon so that the $g$-factor is approximately constant close to $g_0=2$ at low densities \cite{kravchenko2004metal,shashkin2005metal,melnikov2017indication}, this indicates that the renormalized density of states in a spin subband is approximately constant below the Fermi level, independent of the magnetic field. Therefore, the change of $T_{\text F}$ when lifting the spin degeneracy should be controlled by the change of $g_{\text s}$. As concerns the band flattening corresponding to a peak in the density of states at the Fermi level, observed in the 2D electron system in SiGe/Si/SiGe quantum wells, the Fermi energy is not particularly sensitive to this flattening, at least, not too close to the critical point \cite{melnikov2017indication}. So, one expects that the relation $T_{\text F}(0)\approx T_{\text F}(B^*)/2$ holds for the data in question. We stress that its accuracy is not crucial for our qualitative results.

In Fig.~\ref{fig3}, we plot the ratio $\delta\rho/\delta\rho_{\text{max}}=(\rho(T)-\rho(0))/(\rho(T_{\text{max}})-\rho(0))$ as a function of $T/T_{\text{max}}$ in $B=B^*$ so as to check the applicability of the DMFT. The curve for the highest electron density follows the theoretical dependence in the weak-disorder limit at all temperatures. Two other curves for lower electron densities also follow the theoretical dependence at $T\leq T_{\text{max}}$ but deviate from the theory at higher temperatures, revealing the behaviour similar to that observed at low $n_{\text s}$ in zero magnetic field \cite{shashkin2020manifestation}. Albeit the density range of the applicability of DMFT to the completely spin-polarized system is not as wide as that in $B=0$, the low-temperature resistivity drop is described by the theory, similar to the case of the spin-unpolarized electron system. For completeness, in the inset to Fig.~\ref{fig3} we plot the ratio $\rho/\rho_{\text{max}}$ in the fully spin-polarized system as a function of $\rho_{\text{max}}(e^2/\pi h)\,{\text{ln}}(T/T_{\text{max}})$, which is the scaling form suggested by the renormalization-group scaling theory \cite{punnoose2001dilute,punnoose2005metal}. The data do not scale in the range of electron densities studied.

The dynamical mean-field theory successfully describes the closeness of $T_{\text{max}}$ and the renormalized Fermi temperature $T_{\text F}$ in zero magnetic field, as well as the resistivity drop at temperatures below $T_{\text{max}}$ in both spin-unpolarized and fully spin-polarized electron systems. However, the observed decrease of $T_{\text{max}}$ when lifting the spin degeneracy is opposite to the predictions of DMFT. At the same time, the reduced value of $T_{\text{max}}$ in spin-polarizing magnetic fields is consistent with the predictions of the renormalization-group scaling theory, but $T_{\text{max}}$ in zero magnetic field is in disagreement with this theory. The observed behaviour of $T_{\text{max}}$ cannot be described by existing theories.

In view of the competition between electron-electron interactions and disorder, one can expect the value $T_{\text{max}}$ to decrease with increasing disorder level or decreasing electron interaction strength, as occurs in our case when lifting the spin degeneracy. However, the increase of the residual disorder potential when lifting the spin degeneracy in SiGe/Si/SiGe quantum wells, manifested in the metallic regime by the resistivity increase by a factor of a few (the inset to Fig.~\ref{fig1}(b)), cannot be the origin for the observed weakening of the resistivity drop at temperatures below $T_{\text{max}}$ and shift of the maximum in $\rho(T)$ to lower temperatures. Indeed, we compare to the electron system of clean Si MOSFETs, which is different from the studied SiGe/Si/SiGe electron system mainly by lower electron mobility, whereas at the $B=0$ metal-insulator transition in both systems, the values of the interaction parameter are similar, $r_{\text s}\approx 20$ (for details, see Ref.~\cite{melnikov2019quantum}); the interaction parameter is defined as the ratio between the Coulomb and (bare) Fermi energies, $r_{\text s}=g_{\text s}g_{\text v}/2(\pi n_{\text s})^{1/2}a_{\text B}$ (where $a_{\text B}$ is the effective Bohr radius in semiconductor).
Importantly, both systems correspond to the clean limit where the electron interactions are dominant over disorder effects, which is the regime we are interested in. In this case, the zero-magnetic-field metal-insulator transition occurs close to the electron density at which the effective mass at the Fermi level tends to diverge \cite{melnikov2019quantum}. In clean Si MOSFETs, where the electron mobility is some two orders of magnitude lower than that in SiGe/Si/SiGe quantum wells studied here, the resistivity drop at $T<T_{\text{max}}$ in zero magnetic field reaches a factor of 7, which is comparable to that in our samples. Also, in clean Si MOSFETs, the positions of the $\rho(T)$ maxima in $B=0$ closely follow the renormalized Fermi temperatures \cite{radonjic2012wigner,dobrosavljevic2017wigner}, which is similar to our finding in $B=0$. Neither can the reduction of the electron interaction strength due to the reduced spin degrees of freedom in spin-polarizing magnetic fields be the origin of the observed behaviour of $T_{\text{max}}$. Indeed, the relation $T_{\text{max}}\sim T_{\text F}$ still holds for low-mobility Si/SiGe quantum wells in zero magnetic field at electron densities above $n_{\text s}\approx 10^{11}$~cm$^{-2}$ \cite{lu2011termination}, which corresponds to the interaction parameter smaller by a factor of a few compared to that in our range of electron densities. This indicates the spin-related origin of the behaviour of $T_{\text{max}}$ observed in our samples.

It is worth noting that the suppression of the metallic regime and the increase of the transition point when lifting the spin degeneracy in a strongly correlated 2D electron system (see, \textit{e.g}., Fig.~\ref{fig2}, as well as earlier experimental data \cite{dolgopolov1992properties,shashkin2001metal,eng2002effects,jaroszynski2004magnetic}) are explained taking account of the spin and valley degrees of freedom, according to theories \cite{finkelstein1983influence,lee1985disordered,punnoose2001dilute,punnoose2005metal,fleury2008many,dolgopolov2017spin}. Similarly, within the DMFT, the critical density is predicted to increase in spin-polarizing magnetic fields \cite{camjayi2008coulomb}.

In conclusion, we have studied the non-monotonic temperature dependence of the resistivity on the metallic side near the metal-insulator transition in a strongly interacting, spin-unpolarized and spinless two-valley 2D electron system in ultra-clean SiGe/Si/SiGe quantum wells. We have found that in zero magnetic field, the resistivity maximum temperature $T_{\text{max}}$ is close to the renormalized Fermi temperature $T_{\text F}$, which is in agreement with the dynamical mean-field theory. However, rather than increasing along with the Fermi temperature, the value $T_{\text{max}}$ decreases appreciably for spinless electrons in spin-polarizing magnetic fields, which is in contradiction with the theory. The DMFT quantitatively describes the low-temperature resistivity drop in both spin-unpolarized and spinless electron systems. At the same time, the reduced value of $T_{\text{max}}$ in spin-polarizing magnetic fields is consistent with the predictions of the renormalization-group scaling theory, but $T_{\text{max}}$ in zero magnetic field is in disagreement with this theory. Thus, the observed behaviour of $T_{\text{max}}$ cannot be described by existing theories. Nor can it be explained in terms of the increase of the residual disorder potential and the reduction of the electron interaction strength due to the reduced spin degrees of freedom in spin-polarizing magnetic fields, because the relation $T_{\text{max}}\sim T_{\text F}$ still holds for clean Si MOSFETs and low-mobility Si/SiGe quantum wells in zero magnetic field, indicating the spin-related origin of the effect. To describe the observed behaviour of $T_{\text{max}}$ within a single approach that treats properly the spin effects on the low-temperature resistivity in a strongly interacting 2D electron system in parallel magnetic fields, further theoretical efforts are necessary.

\section*{Methods}

The samples we used were ultra-high mobility SiGe/Si/SiGe quantum wells. The peak electron mobility in these samples reaches 240~m$^2$/Vs. The approximately 15~nm wide silicon (001) quantum well was sandwiched between Si$_{0.8}$Ge$_{0.2}$ potential barriers. The samples were patterned in Hall-bar shapes with the distance between the potential probes of 150~$\mu$m and width of 50~$\mu$m using standard photo-lithography. The long side of the Hall bar corresponded to the direction of current parallel to the [110] or [$\bar1$10] crystallographic axis. The in-plane magnetic field was perpendicular to the current to exclude the influence of ridges on the quantum well surface on the resistance measured at high electron densities (for more details, see Refs.~\cite{melnikov2015ultra,melnikov2017unusual}). Measurements were carried out in an Oxford TLM-400 dilution refrigerator with a base temperature of $\approx 25$~mK. Data were taken by a standard four-terminal lock-in technique in a frequency range 0.5--11~Hz in the linear regime of response.

\subsection*{Data availability.} The data that support the findings of this study are available from the corresponding author upon reasonable request.


\subsection*{Acknowledgements}

\noindent The ISSP group was supported by RSF Grant No.\ 22-22-00333. M.M.R.\ acknowledges the funding provided by the Institute of Physics Belgrade through the grant by the Ministry of Education, Science, and Technological Development of the Republic of Serbia.  Numerical simulations were run on the PARADOX supercomputing facility at the Scientific Computing Laboratory of the Institute of Physics Belgrade. The work in Florida was supported by NSF Grant No.\ 1822258 and the National High Magnetic Field Laboratory through the NSF Cooperative Agreement No.\ 1157490 and the State of Florida.  The NTU group acknowledges support by the Ministry of Science and Technology, Taiwan (Projects No.\ 110-2622-8-002-014 and No.\ 110-2634-F-009-027). The Northeastern group was supported by NSF Grant No.\ 1904024.\\
\noindent The authors declare no competing financial or non-financial interests.\\
\noindent Correspondence and requests for materials should be addressed to S.V.K.\\ (email: s.kravchenko@northeastern.edu).

\subsection*{Author contributions statement}

This project was conceived, planned and executed by M.Yu.M., A.A.S., V.T.D.\ and S.V.K. Data were taken by M.Yu.M. and A.A.S. The SiGe/Si/SiGe wafers were grown by S.-H.H. and C.W.L. and processed by M.Yu.M. and A.A.S.  The data analysis was made by M.Yu.M., A.A.S., V.T.D., A.Y.X.Z.\ and S.V.K.  The manuscript was composed by A.A.S. and S.V.K.  All the authors reviewed the manuscript.

\end{document}